\documentclass[twocolumn,english,aps,floatfix,superscriptaddress,showpacs,amsfonts,amssymb,superscriptaddress]{revtex4}
\usepackage{color}

\usepackage{graphicx}
\usepackage{amsmath}
\usepackage{latexsym}
\usepackage{epstopdf}
\usepackage{amsmath}
\usepackage{amssymb,amsmath}
\usepackage{multirow}
\usepackage{mathtools}
\newcommand{\beq}{\begin{equation}}
\newcommand{\eeq}{\end{equation}}
\newcommand{\tn}{\textbf}

\def\ket#1{|#1\rangle}

\begin{document}

\title{Universal behavior of the Shannon mutual information of critical quantum chains}
\author{F.~C.~Alcaraz}

\affiliation{ Instituto de F\'{\i}sica de S\~{a}o Carlos, Universidade de S\~{a}o Paulo, Caixa Postal 369, 13560-970, S\~{a}o Carlos, SP, Brazil}

\author{M.~A.~Rajabpour}
\affiliation{ Instituto de F\'{\i}sica de S\~{a}o Carlos, Universidade de S\~{a}o Paulo, Caixa Postal 369, 13560-970, S\~{a}o Carlos, SP, Brazil}

\date{\today{}}

\begin{abstract}
We consider the Shannon mutual information of subsystems of critical
quantum chains in their ground states.
Our results indicate 
a universal  leading
behavior for large  subsystem sizes. Moreover, as happens with the
entanglement entropy,  its
 finite-size behavior yields  the conformal anomaly $c$ of the underlying
conformal
field theory governing the long distance physics of the quantum chain. We
studied  analytically  a chain of coupled harmonic oscillators and
numerically the $Q$-state Potts models ($Q=2,3$ and $4$), the XXZ quantum chain
and the spin-1 Fateev-Zamolodchikov model. The Shannon mutual information
is a quantity easily computed, and our results indicate that for
relatively small lattice sizes its finite-size behavior already
detects the universality class of quantum critical behavior.
\end{abstract}
\maketitle
Entanglement measures have emerged nowadays as powerful tools for the
study of quantum many body systems\cite{Amico2008,Modi2012}. In one
dimension, where most  quantum critical systems have their
long-distance physics ruled by a conformal field theory (CFT), the
entanglement entropy has been proved  the most important measure
of entanglement. It allows one to identify the distinct universality
classes of critical behaviors. Let us consider a periodic quantum
chain with $L$ sites, and partition the system into  subsystems
{$\cal A$} and {$\cal B$} of length  $\ell$ and $L-\ell$,
respectively. The entanglement entropy is defined as the von Neumann
entropy of the reduced density matrix $\rho_{\cal A}$ of the
partition {$\cal A$}: $S_{\ell} = -Tr_{\cal A} \rho_{\cal A} \ln
\rho_{\cal A}$. If the system is critical and in the ground state,
in the regime where the subsystems are large compared with the lattice spacing, $S_{\ell}$  is given by \cite{Holzey1994,Calabrese2004}
\begin{equation} \label{e1}
S_{\ell} = \frac{c}{3}\ln \left(\frac{L}{\pi}\sin(\frac{\pi\ell}{L})\right) + \gamma_S,
\end{equation}
where $c$ is the central charge of the underlying CFT and $\gamma_s$ is
a non-universal constant. A remarkable fact is that even in the case where the
system is in a pure state formed by an excited state, 
the conformal anomaly dictates the overall behavior of the entanglement, 
similarly as in (\ref{e1}) \cite{excited}.  It is worth mentioning that recently many interesting
methods were proposed \cite{Cardy2011,Demler2012,Zoller2012}  to calculate the entanglement entropy and ultimately central charge, 
however, up to know they have not been implemented
experimentally.
A natural question concerns the possible existence of other measures of
shared information that, similarly as the entanglement entropy, are also
able to detect the several universality classes of critical behavior of quantum critical chains.

In this Letter we present results that indicate  that the
Shannon mutual information of local observables is such a measure. The Shannon mutual
information of the subsystems {$\cal A$} and ${\cal B}$, of sizes $\ell$ and $L-\ell$ is defined as
\begin{equation} \label{e2}
I({\cal A},{\cal B})\equiv Sh({\cal A})+Sh({\cal B})-Sh({\cal A}\cup {\cal B}),
\end{equation}
where $Sh({\cal X}) =-\sum_x p_x \ln p_x$ is the Shannon
entropy of the subsystem ${\cal X}$ with probabilities $p_x$ of being in a
configuration $x$.
These probabilities, in the case where  {$\cal A$} is a  subsystem  of a quantum
chain with wavefunction 
$\ket{\Psi_{{\cal A} \cup {\cal B}}} = \sum_{n,m} c_{n,m}\ket{\phi_{\cal A}^n} 
\otimes \ket{\phi_{\cal B}^m}$, are given by the marginal probabilities 
$p_{\ket{\phi_A^n}} = \sum_m |c_{n,m}|^2$  of the subsystem ${\cal A}$, where 
$\{\ket{\phi_{\cal A}^n}\}$ and $\{\ket{\phi_{\cal B}}^m\}$ 
are the vector basis in subspaces $\cal A$ and $\cal B$.  
 
It is important to notice that the Shannon entropy and the Shannon mutual information are basis dependent
quantities, reflecting the several kinds of observables we can evaluate in the
system and subsystems. Since we are interested  in the evaluation of local
observables that can be measured in any of the subsystems (sizes $\ell =1,\ldots,L$) we consider only  vector basis obtained from the tensor product
$\{\ket{\phi_1}\otimes \ket{\phi_2}\otimes \cdots \}$ of the local spin basis
$\{\ket{\phi_i}\}$ spanning the Hilbert space associated to the site $i$.
In \cite{Wolf2008} it was
conjectured that the mutual information, like the entanglement entropy
should  follow the area law; see also \cite{Bernigau2013}.  Many authors studied the Shannon entropy
of the one dimensional quantum spin chains \cite{Stephan2009,Stephan2010,Zaletel2011} and found useful applications in classifying  one and two dimensional
quantum critical points.
 Several authors also studied
different properties of the Shannon mutual information in two dimensional classical systems \cite{Wilms2011,Wai2013}. They found    that although
the mutual information of two halves of a cylinder has its maximum value at a temperature higher than the critical temperature, its
  derivative  diverges at the critical temperature. Most recently Um et al \cite{Um2012} studied the mutual information
of a subregion with respect to the rest in the periodic transverse
Ising model and surprisingly found that it has the same dependence
$\ln(\sin(\pi \ell/L))$ as (\ref{e1}) but with a distinct
multiplicative constant. 
In this Letter we  study the Shannon mutual
information of local observables in different critical spin chains and argue about their
possible connections to the central charge of the underlying CFT.
 Our results based on the study of several quantum chains
suggest  that for periodic chains in the ground state, the Shannon
mutual information in the scaling regime ($\ell, L >>1$) is
universal and has a  dependence with the subsystem size $\ell$
similar as the one of the entanglement entropy, i. e.,
\begin{equation} \label{e3}
I(\ell,L)= \frac{c} {4}\ln \left(\frac{L}{\pi}\sin(\frac{\pi\ell}{L})\right) + \gamma_I,
\end{equation}
where $c$ is the central charge of the underlying CFT and $\gamma_I$ is a
non-universal constant. Up to the moment, in contrast with the
 entanglement entropy, there is
  no simple general field theoretical method to calculate the Shannon entropy
and consequently the Shannon mutual information. The difficulty
comes from the evaluation of the summation  over the amplitudes  of the
  ground state eigenfunction.

In order to justify our conjecture (\ref{e3}) we calculate the Shannon mutual
information for quantum systems in finite geometries. We first present
our analytical results
  for a system
 of  coupled harmonic oscillators
(Klein-Gordon field theory) and then our numerical analysis for several
critical quantum spin chains:
 $Q$-state Potts model ($Q=2,3$,and $4$), spin-$\frac{1}{2}$ XXZ spin chain and the spin-1 Fateev-Zamolodchikov model in  the antiferromagnetic and ferromagnetic regime.

{\tn {  Harmonic oscillator -}
The Hamiltonian of $L$ coupled harmonic oscillators with
coordinates $\Phi_1,\ldots,\Phi_L$ and conjugated momenta
$\pi_1,\ldots,\pi_L$  is given by
\begin{equation}\label{harmonicOsc}
\mathcal{H}=\frac{1}{2}\sum_{n=1}^{L}\pi_n^2+\frac{1}{2} \sum_{n,n^\prime=1}^L \phi_{n} K_{nn^\prime}\phi_{n^\prime},
\end{equation}
where in the case of nearest-neighbor  couplings the interaction $K$ matrix
is just the discrete Laplacian. In the continuum limit the above Hamiltonian is the one of a
simple scalar free field theory (central charge $c=1$).
Let us now consider $\Phi {_{\cal A}} = (\phi_1, \phi_2, \dots, \phi_{\ell})$ and $\Phi_B = (\phi_{\ell+1}, \phi_{\ell+2}, \dots, \phi_L)$
as the position vectors of the subsystems $\cal A$ and $\cal B$ and $\Pi_{\cal A,\cal B}$
the respective momentum vectors. The Shannon
 mutual information 
$I(\cal A,\cal B)\equiv$$ I(\ell,L)$  between two
regions $\cal A$ and $\cal B$ is
\begin{eqnarray}\label{mutual1}
I(\ell ,L) = \int d^L\Phi p(\Phi_{\cal A},\Phi_{\cal B})\ln \frac{p(\Phi_{\cal A},\Phi_{\cal B})}{p_1(\Phi_{\cal A})p_2(\Phi_{\cal B})},
\end{eqnarray}
where $p(\Phi_{\cal A},\Phi_{\cal B})=\vert \Psi_0\vert ^2$ is the total
and $p_{1}(\Phi_{\cal A})=\int \left[\prod_{m\in {\cal B}}d\phi_m\right]
\vert \Psi_0\vert^2$ and $p_{2}(\Phi_{\cal B})=\int \left[\prod_{m\in ({A})}d\phi_m\right]
\vert \Psi_0\vert^2$ are the reduced probability densities in position 
space ($\Psi_0(\phi_1,\ldots,\phi_L)$ is the ground state
wave function). Then after simple integrations one  get \cite{Unanyan2005}
%
$I(\ell,L)=  \sum_{i=1}^{\ell} \ln (2\nu_i)$,
where  $\nu_i$ are the eigenvalues of
the matrix $C = \sqrt{X_{\cal A} P_{\cal A}}$ and $X_{\cal A}$
and $P_{\cal A}$ are $\ell\times \ell$ matrices
 describing correlations of position and momentum within subsystem $\cal A$ 
\cite{Casini2009}. In other words for $0<i,j<\ell+1$ we have $(X_{\cal A})_{ij}:=<\phi_i\phi_j>=\frac{1}{2}(K^{-\frac{1}{2}})_{ij}$
and $(P_{\cal A})_{ij}:=<\pi_i\pi_j>=\frac{1}{2}(K^{\frac{1}{2}})_{ij}$.
We noticed that the above formula is exactly equal
to the quantum R\'enyi  entanglement entropy with $n=2$ \cite{Casini2009} and consequently using the CFT
techniques \cite{Calabrese2004} one can get the following result for
a periodic system
\begin{eqnarray}\label{mutual0}
I(\ell,L)=\frac{1}{4}\ln(\frac{L}{\pi}\sin(
\frac{\pi \ell}{L})) + \gamma_I,
\end{eqnarray}
that agrees with the conjecture (\ref{e3}).
It is worth mentioning that the mutual Shannon information
$I(\cal A,\cal B)$ obtained in momentum basis also follows the same formula.

\begin{figure}
\begin{center}
\includegraphics[clip,width=0.6\linewidth]{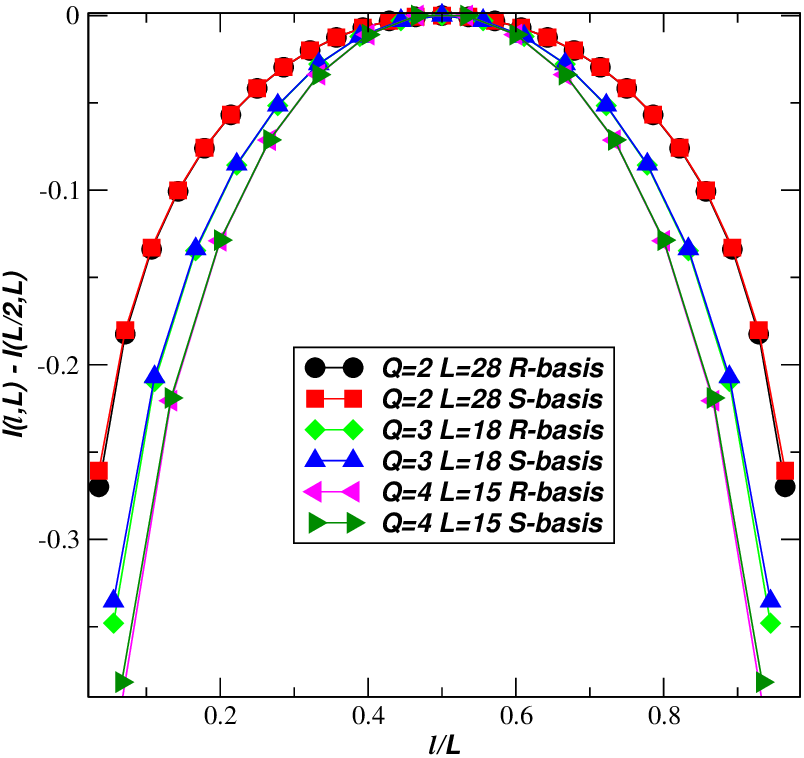}\\
\caption{\label{fig.1} (Color online)
$I(\ell,L)-I(\frac{L}{2},L)$ as a function of the subsystem size for the $Q=2,3$ and $4$ state Potts models in the  $R_i$- and $S_i$-basis.
}
\end{center}
\end{figure}

 \begin{figure}
\begin{center}
\includegraphics[clip,width=0.6\linewidth]{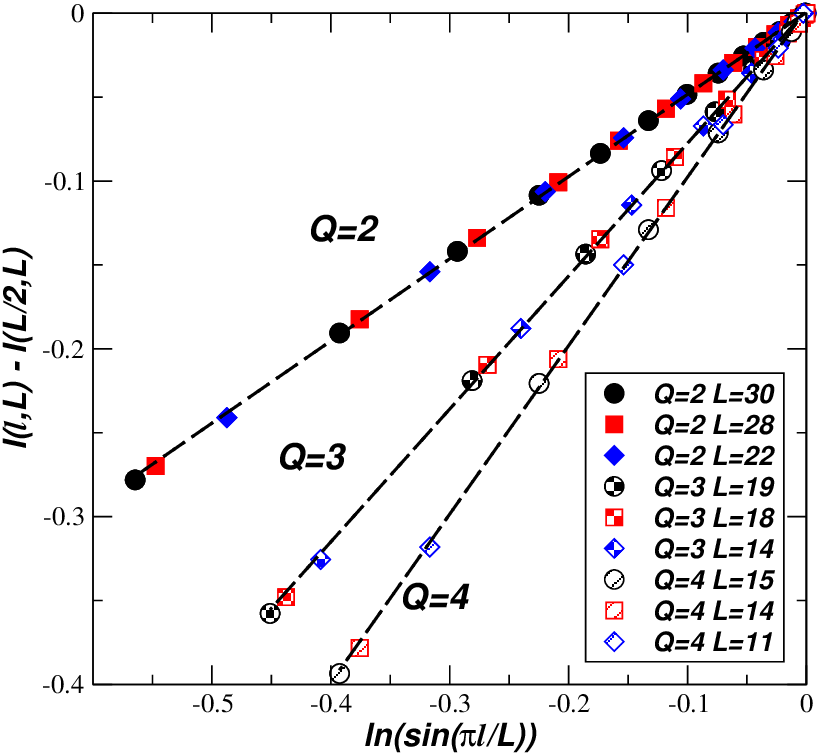}\\
\caption{\label{fig.2} (Color online)
Mutual information 
 for the $Q=2,3$ and $4$ Potts models with different lattice sizes
$L$. The dashed straight lines are obtained from the    fitting by using   the 
largest 
lattice size  shown for each model.(The ground states were taken in the $R_i$-basis). }
\end{center}
\end{figure}

{\tn {Quantum Q-state Potts model -}
The model in a periodic lattice is defined by the Hamiltonian\cite{Wu1982}

\begin{eqnarray}\label{Potts Hamiltonian}
H_Q=-\sum_{i=1}^L\sum_{k=0}^{Q-1}(S_i^kS_{i+1}^{Q-k}+\lambda R_i^k),
\end{eqnarray}
where $S_i$ and $R_i$ are $Q\times Q$ matrices satisfying the following $Z(Q)$ algebra: $[R_i,R_j]=[S_i,S_j]=[S_i,R_j]=0$ for $i\neq j$ and
$S_jR_j=e^{i\frac{2\pi}{Q}}R_jS_j$ and $R_i^Q=S_i^Q=1$.
The system is critical  at the self dual point $\lambda=1$. For $Q=2,3$ and 4 its critical behavior is governed by a CFT with
 central charge $c=1-\frac{6}{m(m+1)}$ where
$\sqrt{Q}=2\cos\frac{\pi}{m+1}$. We first calculate the ground state
of the Hamiltonian (\ref{Potts Hamiltonian}) for $Q=2,3$ and 4 in
different local spin basis by exact diagonalization. 
We verified
for  the critical chains we studied that, as the lattice size
increases, the Shannon  mutual information exhibits a dominant
 behavior, 
 as conjectured in (\ref{e3}). In order to illustrate this
result we show in Fig.~1 the difference $I(\ell,L)- I(L/2,L)$
obtained from the ground-state at the critical point of the $Q=2,3$
and 4 Potts models. The calculations were done by expressing the
ground state in the basis where either the matrices  $R_i$ or $S_i$
are diagonal. Apart from the initial point $\ell=1$ we already see
for these lattice sizes a quite good agreement among the results in
both basis. In order to show the dominant $\ell$-dependence of
$I(\ell,L)$ and test (\ref{e3}) we consider the difference
$I(\ell,L)-I(L/2,L)$, since in this case the 
non universal constant $\gamma_I$ is canceled. In Fig.~2 we plot this difference,
as a function of $\ln(\sin(\pi{\ell}/L)/4$, for some lattice sizes
of the $Q=2,3$ and 4 state Potts models. Clearly the data of
distinct lattice sizes collapse in a straight line in agreement with
(\ref{e3}). The angular coefficient of these lines gives us the
estimate $c(L)$, for  lattice size $L$.
\begin{table}[htp]
\caption{ Numerical values of the constant $c(L)$ for $Q=2,3,4$
Potts models, the XXZ model (XXZ$_{\Delta}$) and the Fateev-Zamolodchikov (FZ$^{\epsilon}_{\gamma}$) model. The expected values for the conformal anomalies
together with  the  lattice sizes used in the  numerical calculation are
also shown. }

\label{tab1}
\begin{tabular}{lccccr}\hline\hline

          &$Q=2$  &    $Q=3$      &    $Q=4$     & XXZ$_{-1/2}$ & XXZ$_{1/2}$\\
\hline
$c$& $\frac{1}{2}$& $\frac{4}{5}$ &    $1$       &       $1$ & $1$    \\
$c(L)$&    0.49     &  0.79        &1.00         &1.00             &1.03  \\
$L$&     30      &   19           &   14          &     30       &30 \\
 \hline\hline
          & XXZ$_{0}$  & FZ$_{\pi/3}^1$  &  FZ$_{\pi/4}^1$ &   FZ$_{\pi/3}^{-1}$   &    FZ$_{\pi/3}^{-1}$\\
\hline
$c$&     1   &   $\frac{3}{2}$&    $\frac{3}{2}$ &     $1$ & $1$\\
$c(L)$&    1.02 &        1.53           &1.47 & 1.03&1.06\\
$L$&     30          &20 & 20 &20      &20 \\
\hline \hline

\end{tabular}
\end{table}
In table 1 we give these estimates.
The agreement with the predicted values
are remarkable already for the lattice sizes we considered.

{\tn {XXZ quantum chain -}}
The model describes the dynamics of spin-$\frac{1}{2}$ particles given by the Hamiltonian
\begin{eqnarray}\label{XXZ}
H_{\text{XXZ}}=-\sum_{i=1}^L(\sigma_j^x\sigma_{j+1}^x+\sigma_j^y\sigma_{j+1}^y+\Delta\sigma_j^z\sigma_{j+1}^z),
\end{eqnarray}
where $\sigma^x,\sigma^y\,\sigma^z$ are spin-$\frac{1}{2}$ Pauli
matrices and $\Delta$ an anisotropy. This model provides an
interesting check for the universal behavior of the Shanon mutual
information. It has  a continuous critical line, $-1 \leq
\Delta <1$, whose CFT has a central charge $c=1$. According to
(\ref{e3}) we should expect a data collapse of $I(\ell,L) -
I(L/2,L)$ for distinct lattice sizes and anisotropies.
 \begin{figure}
\begin{center}
\includegraphics[clip,width=0.6\linewidth]{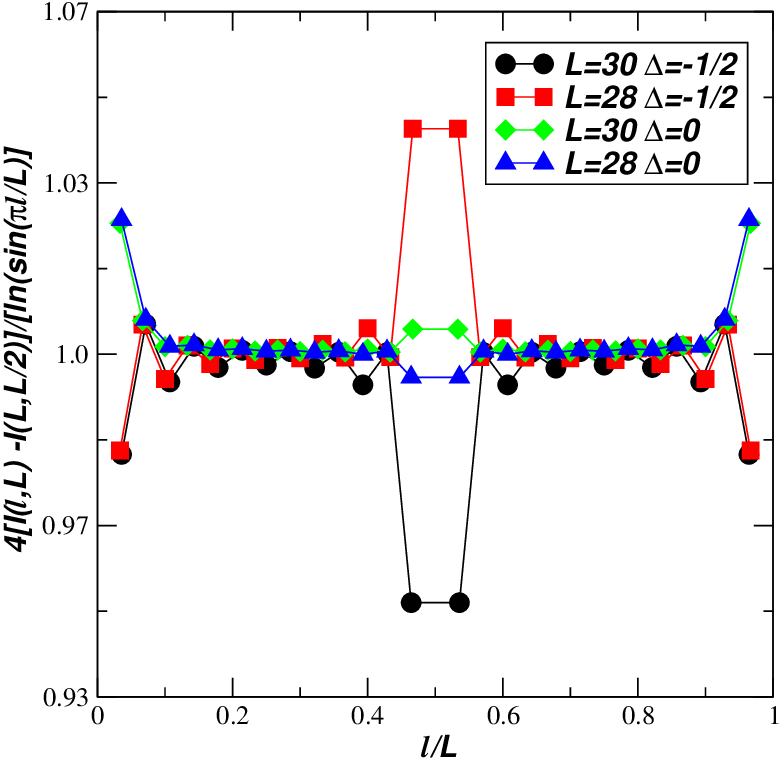}\\
\caption{\label{fig.3} (Color online)
 Mutual information as a function of  the subsystem size for the  XXZ model
with  different values of the anisotropy  $\Delta$ and lattice sizes $L$.}
\end{center}
\end{figure}

In order to illustrate this fact we plot in Fig.~3 the ratio
$[I(\ell,L) - I(L/2,L)]/[\ln(\sin(\pi\ell/L))/4]$ as a function of
${\ell}/L$. We clearly see that for distinct anisotropies and
lattice sizes the ratio is close \cite{foot1} to a constant value
given by the conformal anomaly. Similarly as we did for the Potts
models the estimated values $c(L)$ for the anisotropies $\Delta=0$
and $\Delta=\pm\frac{1}{2}$ are shown in table 1.

{\tn {Fateev-Zamolodchikov model  -}
This is a spin 1 model whose Hamiltonian is given by \cite{fateev1}
\begin{eqnarray} \label{FZ}
H_{FZ} &=& \epsilon \sum_{i=1}^{L} \{ \sigma_i - (\sigma_i^z)^2
-2(\cos \gamma -1)(\sigma_i^{\perp}\sigma_{i}^z +
\sigma_i^z\sigma_{i}^{\perp} \nonumber \\ &&-2\sin^2\gamma (\sigma_i^z
-(\sigma_i^z)^2 + 2(S_i^2)^2\},
\end{eqnarray}
where $\vec{S} = (S^x,S^y,S^z)$ are spin-1 $SU(2)$  matrices,
$\sigma_i^z=S_i^zS_{i+1}^z$ and
 $\sigma_i=\vec{S}_i\vec{S}_{i+1} = \sigma_i^{\perp} +\sigma_i^z$.
The model is antiferromagnetic  for $\epsilon =+1$ and ferromagnetic
for $\epsilon =-1$.
This is an important check for the conjecture (\ref{e3}) since the model
 has a line of critical points
 ($0\leq\gamma \leq \frac{\pi}{2}$) with a quite distinct behavior in the
 antiferromagnetic ($\epsilon =+1$) and ferromagnetic
($\epsilon=-1$) cases.
The antiferromagnetic version of the model  is
governed by a CFT with central charge $c=\frac{3}{2}$ \cite{fateev2} while the ferromagnetic
one   is ruled by a $c=1$ CFT \cite{fateev3}.
\begin{figure}
\begin{center}
\includegraphics[clip,width=0.6\linewidth]{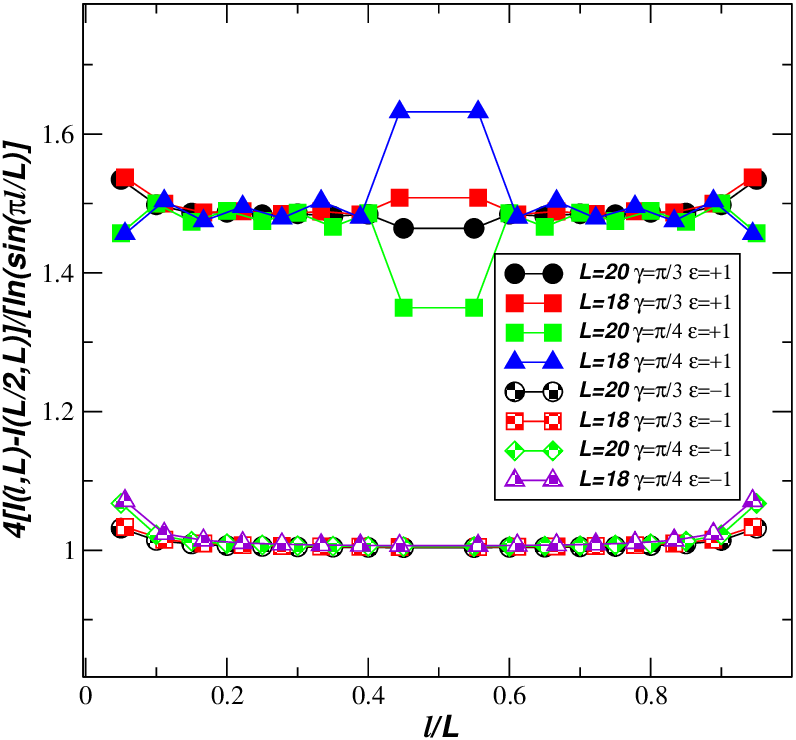}\\
\caption{\label{fig.4} (Color online)
 Mutual information dependence with the subsystem size for the   Fateev-Zamolodchikov model with different values of $\gamma$ and
lattice  sizes.}
\end{center}
\end{figure}

In Fig.~4 we show for $\epsilon =\pm1$, the ratio
$[I(\ell,L)- I(L/2,L)]/[\ln(\sin(\pi\ell/L))/4]$ as a function of
$\ell/L$.
The data are  shown for the anisotropies $\gamma=\pi/3$ and $\gamma=\pi/4$
and for lattice sizes $L=18$ and $L=20$. We clearly see an agreement
with the expected central charge of the corresponding CFT. In table 1 we
give the predicted values $c(L)$,
 and the agreement is again quite good.

 \begin{figure}
\begin{center}
\includegraphics[clip,width=0.6\linewidth]{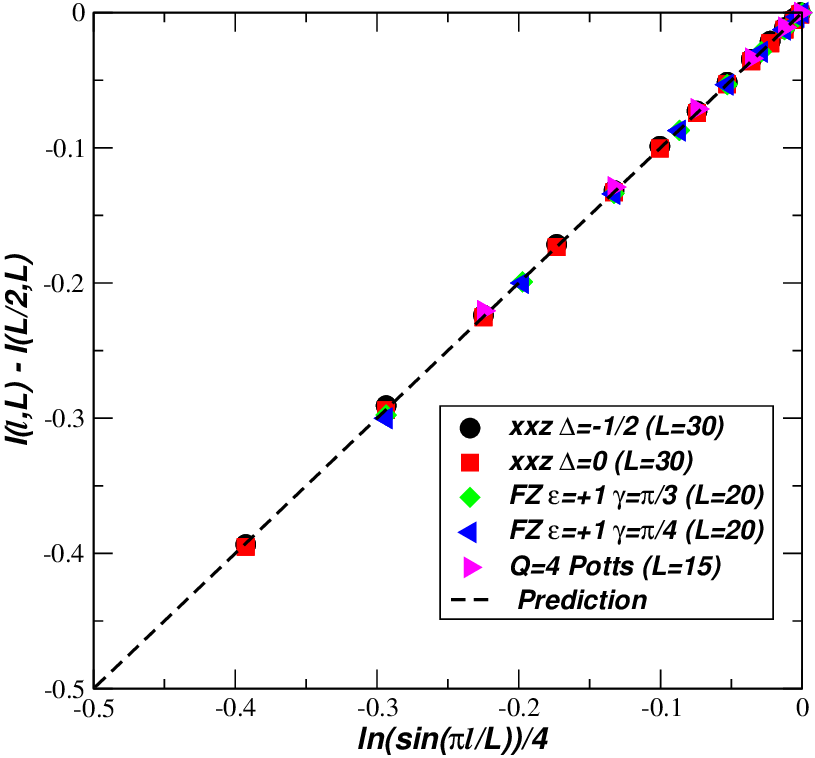}\\
\caption{\label{fig.5} (Color online)
Mutual information for some lattice sizes for the
 XXZ chain with anisotropies $\Delta=-1/2$ and $\Delta=0$, the ferromagnetic
spin-1 Fateev-Zamolodchikov model with anisotropies
$\gamma=\pi/3$ and $\gamma=\pi/4$ and for the $4$-state Potts model. All these models are ruled by a CFT
with conformal anomaly $c=1$.}
\end{center}
\end{figure}

{\tn {Conclusions -} All the analytical and numerical calculations
presented in this Letter indicate the following properties for the
Shannon mutual information of local observables for the ground state of critical quantum
chains: a) The leading dependence with the subsystem size $\ell$ 
characterizes the  universality class of the
critical behavior of the quantum chain, b)  The finite-size scaling
function $\ln(\frac{L}{\pi}\sin  \frac{\pi{\ell}}{L})$ is the same
as that of the entanglement entropy, c) The finite-size scaling,
similarly to the entanglement entropy, is proportional to the central
charge of the underlying CFT.  An overall illustration of these
points is presented in Fig.~5, where we show for $2\leq\ell <L-1$
the finite-size behavior of the Shannon mutual information for the
different models with central charge $c=1$ presented in this letter.
This figure shows that models whose Hamiltonians acts on rather
distinct Hilbert spaces share the same universal behavior for the
Shannon mutual information of their ground states. Our results 
 indicate that the Shannon mutual
information, similarly as  the entanglement entropy provide  excellent tools
 for the evaluation  of the central charge of conformal 
invariant quantum chains.
Although the numerical results presented in 
this Letter are obtained by using the Lanczos  method, we also 
verified that the Shannon mutual 
information can also 
be computed \cite{xavier}, 
 for relatively large lattices,  by using the DMRG \cite{white}.
%

Finally we mention that the conjecture (\ref{e3}) announced in this paper 
raises several interesting
questions to be answered in the future. The first one concerns its proof 
based on CFT calculations as
done by Calabrese and Cardy \cite{Calabrese2004} in the case of the
entanglement entropy. In the case of the harmonic
oscillator chain  we show that in the bulk limit the dominant part of the
Shannon mutual information  is the same as that of  the $n=2$ R\'enyi 
entanglement entropy. 
 A  proof of this equivalence would  produce  as a corollary the  conjecture
(\ref{e3}). 

This work was supported in part by FAPESP and CNPq (Brazilian agencies). We 
thanks J. A. Hoyos and R. Pereira for useful discussions and a careful reading 
of the manuscript.

%



\begin{thebibliography}{44}
\bibitem{Amico2008}L. Amico,R. Fazio,A. Osterloh, and V. Vedral, Rev. Mod. Phys. {\bf{80}}, 517 (2008). 

\bibitem{Modi2012}  K. Modi, A. Brodutch, H. Cable, T. Paterek, 
and V. Vedral, Rev. Mod. Phys.  {\bf{84}}, 1655 (2012). 

\bibitem{Holzey1994} C. Holzhey, F. Larsen, and F. Wilczek, Nucl. Phys. B 
{\bf 424}, 443 (1994).

\bibitem{Calabrese2004}P. Calabrese and J. Cardy, J. Stat. Mech. (2004)  P06002 .
\bibitem{excited} F. C. Alcaraz, M. I. Berganza, and G. Sierra, Phys. Rev. Lett. {\bf 106}, 201601 (2011);
M. I. Berganza, F. C. Alcaraz, and G. Sierra, J. Stat. Mech. (2012)  P01016; L. Taddia, J. C. Xavier, F. C. Alcaraz, and G. Sierra (unpublished).

\bibitem{Cardy2011} J. Cardy, Phys. Rev. Lett. {\bf{106}}, 150404 (2011). 
\bibitem{Demler2012}D. A. Abanin and E. Demler, Phys. Rev. Lett. {\bf{109}}, 020504 (2012). 
\bibitem{Zoller2012}
A. J. Daley, H. Pichler, J. Schachenmayer, and P. Zoller, Phys. Rev. Lett. {\bf{109}},020505 (2012).



\bibitem{Wolf2008} M. M. Wolf, F. Verstraete, M. B. Hastings, J. I. Cirac, Phys. Rev. Lett. {\bf{100}}, 070502 (2008). 

\bibitem{Bernigau2013} H. Bernigau, M. J. Kastoryano, and J. Eisert, [arXiv:1301.5646].

\bibitem{Stephan2009}J-M St\'ephan, S. Furukawa, G. Misguich, and V. Pasquier,
Phys. Rev. B, {\bf{80}}, 184421 (2009). 

\bibitem{Stephan2010} J-M St\'ephan, G. Misguich, and V. Pasquier, Phys. Rev. B, {\bf{82}}, 125455 (2010);
Phys. Rev. B {\bf{84}}, 195128 (2011).

\bibitem{Zaletel2011} M. P. Zaletel, J. H. Bardarson, and J. E. Moore, Phys. 
Rev. Lett.  {\bf{107}},  020402 (2011).


\bibitem{Wilms2011} J. Wilms, M. Troyer, and F. Verstraete, J. Stat. Mech.  (2011) P10011;
J. Wilms, J. Vidal, F. Verstraete,and  S. Dusuel, J. Stat. Mech. (2012)P01023.

\bibitem{Wai2013} H. W. Lau  and P. Grassberger,   Phys. Rev. E  {\bf{87}}, 022128 (2013). 


\bibitem{Um2012} J. Um, H. Park and H. Hinrichsen, J. Stat. Mech. (2012)  P10026 
.
\bibitem{Unanyan2005}R. G. Unanyan and M. Fleischhauer, Phys. Rev. Lett. {\bf{95}}, 260604 (2005). 


\bibitem{Casini2009}H. Casini and M. Huerta, J. Phys. A {\bf{42}}, 504007 (2009).

\bibitem{Wu1982}  F. Y. Wu, Rev. Mod. Phys, {\bf{54}}, 235 (1982). 

\bibitem{foot1}
 The small
deviations that appears for $\ell =L/2-1$ is enlarged due to the  division by
$\ln(\sin(\pi\ell/L))$.

\bibitem{fateev1} A. B. Zamolodchikov and V. A. Fateev, Yad. Fiz. {\bf 32}, 581 (1980) [Sov. J. Nucl. Phys. {\bf 32}, 298 (1980)].

\bibitem{fateev2}
  F. C. Alcaraz and M. J. Martins, Phys. Rev. Lett. {\bf{61}}, 1529 (1988); 
P. Di Francesco, H. Saleur, and J.-B. Zuber, Nucl. Phys.
B {\bf{300}}, 393 (1988).

\bibitem{fateev3} 
F. C. Alcaraz and M. J. Martins, Phys. Rev. Lett. {\bf{63}}, 708 (1989). 


\bibitem{xavier} F. C. Alcaraz, M. Rajabpour, and J. C. Xavier (unpublished).

\bibitem{white} S. R. White, Phys. Rev. Lett. {\bf 69}, 2863 (1992).


\end{thebibliography}
\end{document}